# Investigating the optical properties of carbonado-diamonds

Jozsef Garai[a,*], Stephen E. Haggerty[a], and Andriy Durygin[b]
[a] Dept. of Earth Sciences, Florida International University, PC-344, Miami, FL 33199, USA
[b] CeSMEC, Florida International University, VH-150, Miami, FL 33199, USA
[*]Corresponding author. Tel: 786-247-5414; e-mail: Jozsef.garai@fiu.edu

**Abstract**

Carbonado-diamond is the most enigmatic of diamond types, and surprisingly there are few optical studies. In some cases, the diamond peak at 1332.5 $cm^{-1}$ is absent, and the presence of a D-band graphite peak around 1320 $cm^{-1}$ is observed in our Raman investigation. The disappearance of the diamond peak in carbonado-diamond is interpreted as the combined effect of small crystal size and the high absorption of graphite. The photoluminescence spectra of carbonado-diamond shows two new zero phonon lines at 584.2 nm (2.12 eV) and at 743 nm (1.67 eV). By heating the carbonado-diamond to 1500 °C for ten minutes in argon the appearance of an additional peak at 693.5 nm (1.786 eV) is detected. The infrared absorption spectra of carbonado-diamond powder indicate either extremely low or extremely high absorption.

## 1. Introduction
*1.1 Origin*

Carbonado-diamond is found only in Brazil and the Central African Republic in non-traditional diamond deposits. The mineral is black, porous, and strongly aggregated, with diamond grain sizes in the nano- to micro-meter range [1]. There is no consensus for the origin of carbonado-diamond and models of formation are extremely varied. Crust-affiliated minerals rather than deep mantle mineral inclusions in carbonado-diamond, coupled with light carbon isotopic compositions has led to the suggestion that carbonado-diamond formed by meteoritic impact [2,3]. This was later supported by the inference that the largest magnetic anomaly on Earth, Bangui in the Central African Republic, is a buried iron meteorite [4]. Other interpretations for the origin of carbonado-diamond include ion-implantation by U and Th in coal deposits [5]; formation by a complex series of reactions in the upper mantle and subsequently in the crust [6]; ocean floor subduction [7], followed by mantle plume interaction [8]; or by a combined growth and sintering process at high P & T [9]. In the latest review [10] a comparison of carbonado-diamond is made with loosely bonded framesite from conventional diamond deposits. Although the coverage is comprehensive, neither are sufficiently well understood to reach any new or profound genetic conclusions.

*1.2 Spectral properties*

Photoluminescence spectra of carbonado-diamond exhibit zero-phonon lines at 1.945 and 2.156 eV, have been reproduced in Type 1 diamond that was irradiated and then heated in the range 550-1500°C [11]. A characteristic photoluminescence zero-phonon line is also present at 2.463 eV which is attributed to paired nitrogen atoms in association with a vacancy; this possibly points to relatively long term annealing for carbonado-diamond. These interpretations are supported by photoluminescence and electron paramagnetic resonance measurements by Nadolinny et al. [12], who also report a hydrogen-containing defect, previously only observed in CVD diamonds. Large (~200 μm) and rare mono-crystals of diamond in carbonado show luminescence in shades of orange and tones of green [13].

The presence of a broad band absorption around 1100 cm$^{-1}$, resulting from contaminated silica, prevents the measurement of infrared absorption of carbonado-diamond in the most important band range of diamonds between 1000-1300 cm$^{-1}$ [6]. No experiments have succeeded in overcoming these problems and the type of diamond in carbonado has yet to be identified. In this study the optical properties of carbonado-diamond have been investigated in an attempt to shed light on this mysterious mineral.

## 2. Experimental

*2.1 Raman*

The measurements were conducted at room temperature by using the Raman spectrometer in the back-scattering configuration. Ti$^{3+}$/sapphire laser pumped by an argon ion laser was tuned at the near infrared wavelength of 785 nm, which can significantly suppress the strong fluorescence in diamond. Raman spectra were collected by using a high throughput holographic imaging spectrograph with volume transmission grating, holographic notch filter and thermoelectrically cooled CCD detector with a resolution of 4 cm$^{-1}$.

*2.2 Fluorescence*

The fluorescence experiments were performed at liquid nitrogen temperature in the spectral range of 520-980 nm. An argon ion laser (514 nm) was used for excitation. The emitted light was dispersed in a Holospec VPT System (Kaiser Optical Syst. Inc.) with DV420-OE CCD Detector (Andor Technology).

*2.3 Infrared absorption*

In order to remove silica contamination from the carbonado-diamonds, the samples were crushed to <67 μm. The resulting powders were kept in HF pressure bombs at 60 °C for three days. The recovered carbonado-diamond was mixed with KBr in different proportions and then pressed under low vacuum to form ~1 mm thick pellets. Using the same procedure, pellets with the same weight proportion were made from natural diamond powder crushed to ~25 μm.

The absorption of the pellets were measured using a Nicolet Magna IR-560 high precision Fourier-transform infrared spectrometer with a resolution of up to 0.025 wavenumbers. Using a broadband IR source, and NaCl windows, allowed a study of the 900-4200 $cm^{-1}$ region of the infrared spectrum.

## 3. Results and discussion

*3.1 Raman*

The Raman spectra of three Brazilian (BR), eight Central African (CAR), and one unknown source carbonado-diamonds were investigated. The identical diamond Raman shift peak (1332.5 $cm^{-1}$) is present in most of the samples. The intensity of the peak in the carbonado-diamonds is smaller than in diamond and in some cases is not identified. The disappearance of a sharp diamond peak at 1332.5 $cm^{-1}$ is replaced by the appearance of a D band peak at ~1320 $cm^{-1}$ (Fig. 1a).

Single crystal graphite shows one single line at 1575 $cm^{-1}$ (G-band) [14]. In commercial graphite, activated charcoal, stress-annealed pyrolitic graphite, lampblack, and vitreous carbon, another line at 1355 $cm^{-1}$ has been identified. The 1355 $cm^{-1}$ line is defect-induced and is called D line or D mode. The D-band position varies between 1300–1450 $cm^{-1}$ as a function of excitation energy [15]. Using a 785 nm laser the position of the D-band should be around 1320 $cm^{-1}$. It is suggested that the wide peak at ~1320 $cm^{-1}$ is possibly related to graphite which is present in the carbonado-diamond. Most of the main diamond peak in the carbonado-diamond has a well developed shoulder (Fig. 1b) which can be explained by an overlapping D-band graphite peak. The carbonado-diamond spectra are consistent with the spectra of CVD diamonds with methane concentration of 1.44-2.16% and excited at 780 nm. [15].

The disappearance of the well defined Raman diamond line at ≈1332.5 $cm^{-1}$ could be the result of strong optical absorption of graphite. This peak is not identified in CVD diamond when the methane concentration is higher than 2% [15]. Based on density measurements, diamond produced at 2% methane concentration should contain a 5% graphite [16].

The other explanation for the disappearance of the diamond Raman line might be the small crystal size resulting from the microcrystalline structure of carbonado-diamond. The minimum size "visible" by Raman investigation is between 1-50 nm [17]. From the particle size effect on diamond, it has been suggested that the primary Raman peak of diamond would be difficult to detect when the crystal size is < 5nm [18]. The Raman spectra of the synthetic micro-diamond (3-6 nm) do not show an identifiable diamond Raman shift peak at 1332.5 cm$^{-1}$(Fig. 1c).

The disappearance of the diamond Raman peak in the carbonado-diamond most likely is the combined effect of the small crystal size and the presence of D-band graphite. It is not possible to estimate the individual contribution of these two effects. However, we may conclude that a nanometer size fraction is present in carbonado-diamonds, and coupled with the possible presence of defect induced graphite, that a CVD-like process was operative in the genesis of carbonado-diamonds [19].

*3.2 Fluorescence*

The fluorescence spectra of the same samples were collected. The defect spectra were dominated by zero phonon lines (ZPL) from nitrogen-related defect centers at nominal energies of 1.945 eV (640 nm) and 2.156 eV (575 nm). The intensity of the N-V (1.945 eV) center was higher in all cases than the intensity of the N-V-V (2.154 eV) centers. The latter was even absent in one of the Central African carbonado-diamond (CAR-32). In a previous study [20], that investigated the effect of laser excitation energy between 450-650 nm, on the photoluminescence of diamond films grown by various techniques (HF-CVD, MW-PECVD, DC arc-jet) it was shown that the 514 nm Ar laser, was able to excite the N-V (1.945 eV) centers, while the excitation of the N-V-V (2.154 eV) was absent. The weaker intensity of the N-V-V (2.154 eV) in our experiments is interpreted as the result of the excitation wavelength. The intensity of the fluorescence in carbonado-diamond is usually about one order of magnitude higher in comparison to standard diamonds. Carbonado-diamonds have typically smooth surfaces but porous interiors. The fluorescence spectra on the broken surfaces is almost identical to the original smooth surface, except the intensity is generally higher (Fig. 2).

Two additional zero phonon lines at 584.2 nm (2.12 eV) and 743 nm (1.67 eV) are observed in one Brazilian and two Central African samples (Fig. 3). Neither of these peaks have previously been reported for carbonado-diamond. The 1.67 eV center most likely relates to a neutral vacancy (GR1 centers) [21] but the possibility of chomian impurities [22] can not be excluded. No previous report of 584.2 nm peak in the literature has been found for diamonds [e.g. 23;24]. The newly detected peaks, 584.2 nm (2.12 eV) and 743 nm (1.67 eV) were observed on the same sample, on the original surface, with the exception of CAR-32.

Two carbonado-diamonds, were heated by a $CO_2$ laser to 1500 $^0$C in an argon flow for ten minutes. The intensity of the fluorescence is greatly reduced. In one case, the peak related to N-V-V (2.154 eV) centers disappeared and in both cases an additional peak at 693.5 nm (1.786 eV) is observed.

*3.3 Infrared absorption*

The infrared absorption spectra of the pellets were collected between 1000-1500 $cm^{-1}$. In the natural diamond samples the prominent bands, 1185 $cm^{-1}$ (B-Aggregate) and 1282 $cm^{-1}$ (A-Aggreagte) are clearly present. None of the investigated carbonado-diamond pellets showed any absorption peak (Fig. 4). The lack of absorption peaks in the carbonado-diamonds can be the result of extremely low or extremely high absorption of the carbonado-diamond. The detected presence of graphite in the carbonado-diamond supports an extremely high IR absorption possibly resulting from the presence of graphite.

## 4. Conclusions

The Raman investigations identified graphite D-band peak around 1320 $cm^{-1}$ in the carbonado-diamond. The detected graphite in the carbonado-diamonds a new line of permissive evidence that carbonado-diamond is related to a CVD like process.

The diamond peak at 1332.5 $cm^{-1}$ disappears entirely in some cases. The disappearance of the diamond peak is interpreted as the combined effect of the microcrystalline structure of the carbonado-diamond and the presence of graphite with defects.

The fluorescence investigations detected two new, previously not reported, peaks at 584.2 nm (2.12 eV) and 743 nm (1.67 eV) in the carbonado-diamond spectra. It is very likely the 1.67 eV center relates to a neutral vacancy (GR1 centers). The heat treatment of the carbonado-diamond (1500 $^0$C) resulted in an additional peak at 693.5 nm (1.786 eV).

The carbonado-diamond powder-KBr pellets did not show any absorption peak in the investigated infrared region. The lack of absorption can be explained by either extremely low or extremely high absorption. The identified graphite in the carbonado-diamond suggests that a lack of absorption peaks might result from extremely high absorption.

**Acknowledgement**

We would like to thank to the Department of Chemistry and Biochemistry at Florida International University for providing the equipment and the assistance for the infrared absorption measurements.

**References:**

[1] S.E. Haggerty, Science, 285 (1999) 851.
[2] J.V. Smith, and J.B. Dawson, Geology, 13 (1985) 342.
[3] K. Shibata,.H. Kamioka, F. V. Kaminsky, V. I. Koptil, and D. Svisero, Mineral. Mag., 57 (1993) 607.
[4] R.W. Girdler, P.T. Taylor, and J.J. Frawley, J.J. Tectonophys., 212 (1992) 45.
[5] F.V. Kaminski, Dokl. Acad. Nauk. SSSR, 294 (1987) 439.
[6] H. Kagi, K. Takahashi, H. Hidaka, and A. Masuda, Geochem. Cosmochim. Acta, 58 (1994) 2629.
[7] P.S. DeCarli, EOS Amer. Geophys. Union (1997) S333.
[8] T. Irifune, A. Kurio, S. Sakamoto, T. Inoue, and H. Sumiya Nature, 421 (2003) 599.
[9] J.H. Chen, and G. Van Tendeloo, J. Electon. Microscopy 48 (1999) 121
[10] P.J. Heaney, E.P. Vicenzi, and S. De, Elements (2005) 85.
[11] C.D. Clark, A.T. Collins, and G.S. Woods, In: The Properties of Natural and Synthetic Diamond, ed. J.E. Field. Academic Press, 1992 p. 35.
[12] V.A. Nadolinny, V.S. Shatsky, N.V. Sobolev, D.J. Twitchen, O.P. Yuryeva, I.A. Vasilevsky, and V.N. Lebedev, V.N. Am. Mineral. 88 (2003) 11.
[13] McGee, C.W. & Taylor, W.R., Proc. VIIth Int. Kimberlite Conf. (1999) 529.
[14] F. Tuinstra, and J.L. Koenig, J. Chem. Phys. 53 (1970) 1126.
[15] S. M. Leeds, T. J. Davis, P. W. May, C. D. O. Pickard, and M. N. R. Ashfold Diamond and Rel. Mat. 7 (1998) 233.
[16] Y. Sato, and M. Kamo In: The Properties of Natural and Synthetic Diamond, ed. J.E. Field. Academic Press Limited, London, UK, 1992 p. 439
[17] E. Duval, A. Boukenter, and B. Champagnon B. Phys. Rev. Lett. 56 (1986) 2052.
[18] M. Yoshikawa, Y. Mori, M. Maegawa, G. Katagiri, H. Ishida, and A. Ishitani, Appl. Phys. Lett. 62 (1993) 3114.
[19] S.E. Haggerty, In: Proc. 5th Int. Sym. Adv. Mat., Tsukuba (1998) 39.
[20] M.C. Rossi, S. Salvatori, F. Galluzzi, F. Somma, and R.M. Montereali, Diamond and Related Mat. 7 (1998) 255.
[21] J. Lindblom, J. Hölsä, H. Papunen, and H. Häkkänen Amer. Mineral. 90 (2005) 428.
[22] A.M. Zaitsev, Phys. Rev. B 61 (2000) 12909.
[23] J. Walker, Rep. Prog. Phys. 42 (1979) 1605.
[24] The Properties of Natural and Synthetic Diamond, ed. J.E. Field. Academic Press Limited, London, UK, 1992 p. 687

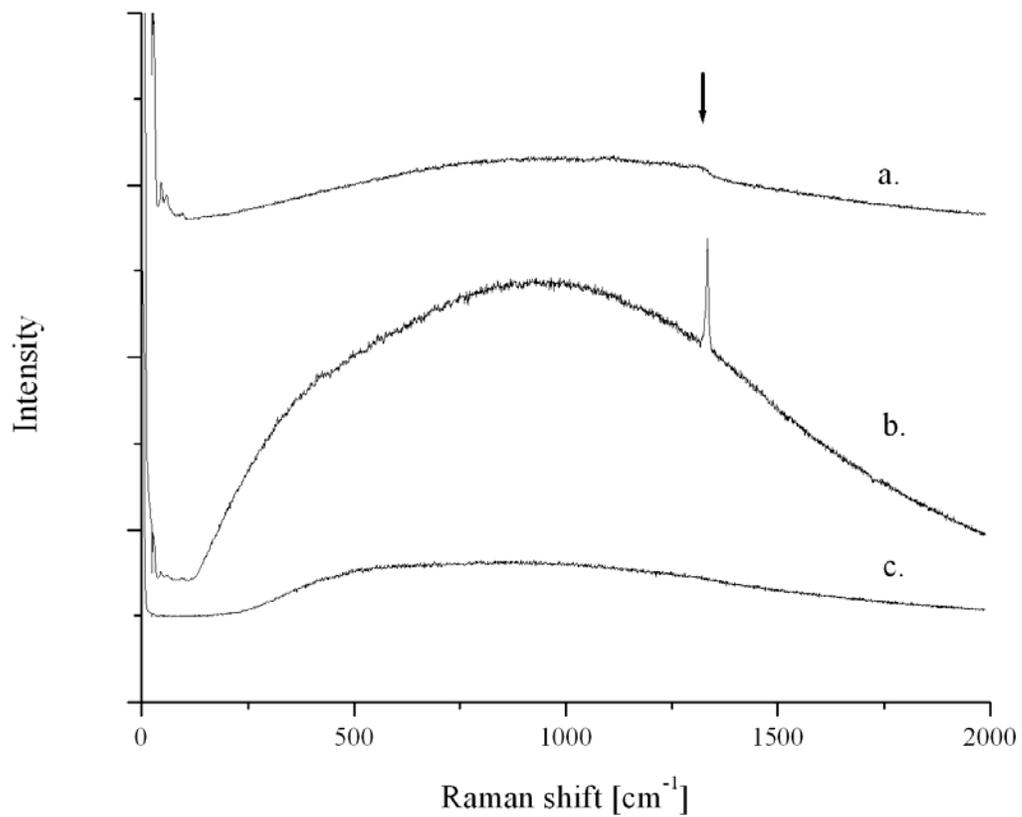

Fig. 1 The Raman spectra of carbonado-diamond. The graphs are shifted for better visibility. a. Total disappearance of the 1332.5 cm$^{-1}$. Incipient development of the graphite D-band is marked with an arrow. b. Regular spectra with an asymmetric diamond peak at 1332.5 cm$^{-1}$. c. Spectra of the 3-6 nm size synthetic micro-diamond.

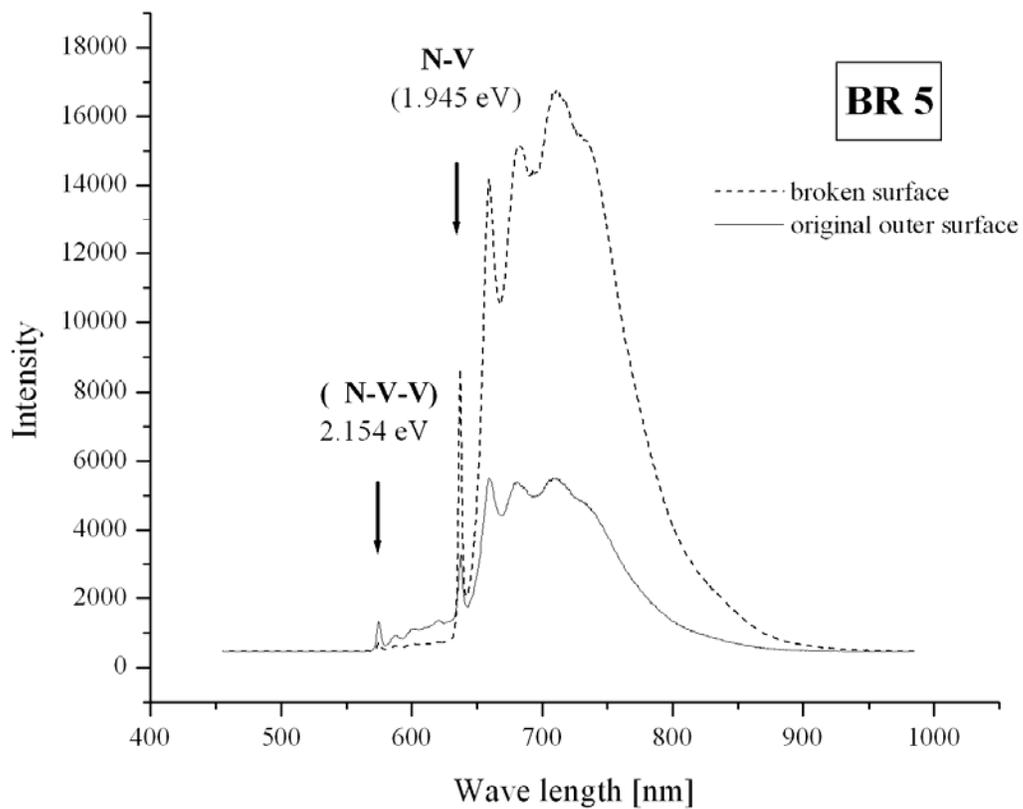

Fig. 2 Representative photoluminescence spectra of carbonado-diamond at 77 K. The spectra was collected for 0.2 sec.

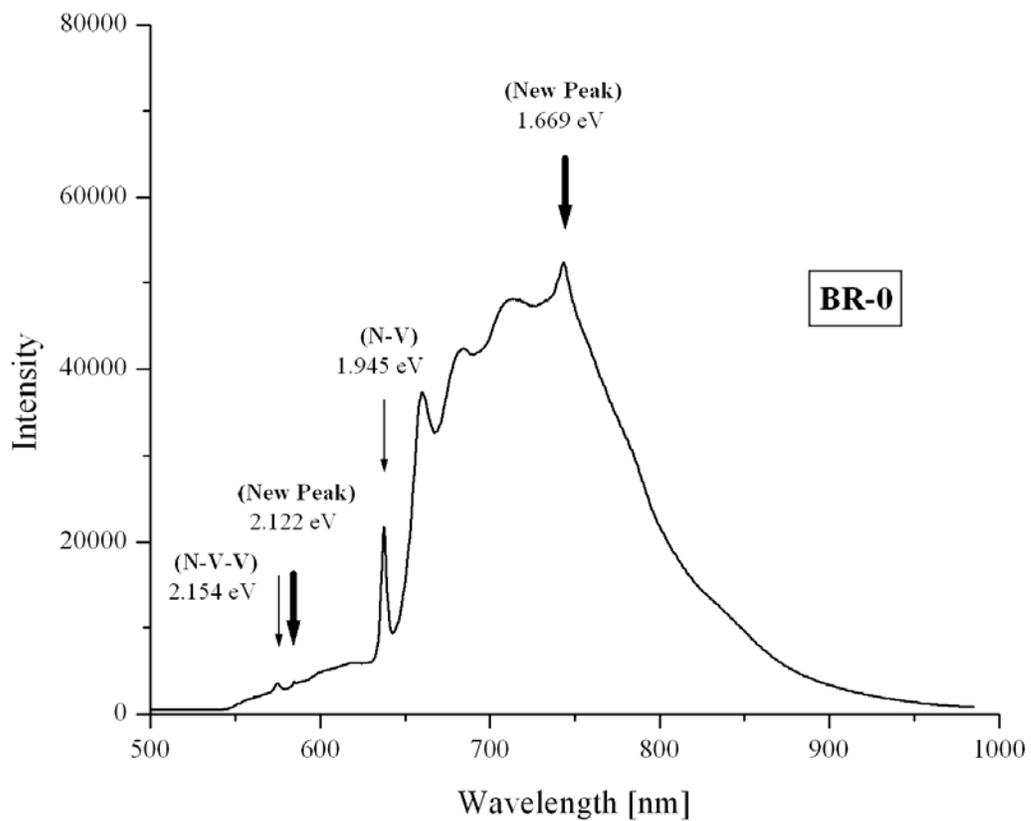

Fig. 3  The photoluminescence spectra of carbonado-diamond showing the two detected new peaks.

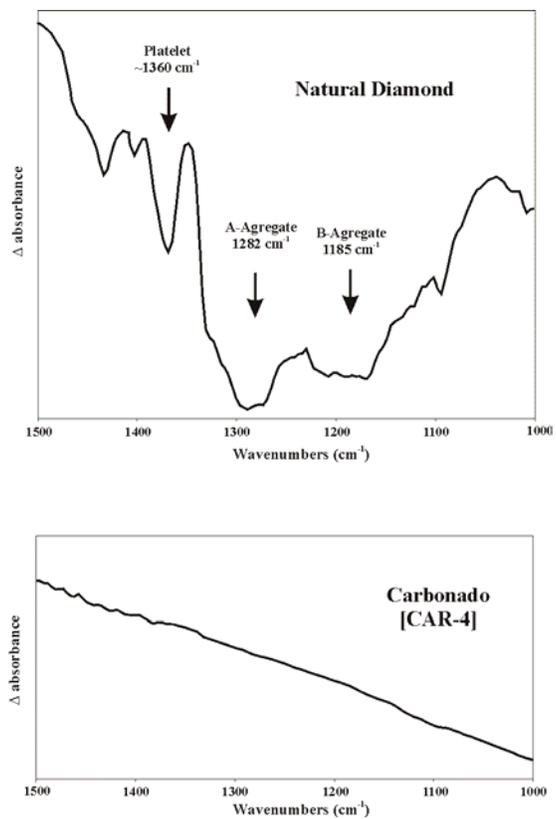

Fig. 4 Infrared absorption spectra of natural diamond and carbonado-diamond powder in KBr.